\documentclass[lettersize,journal]{IEEEtran}
\usepackage{amsmath,amsfonts}
\usepackage{array}
\usepackage{textcomp}
\usepackage{url}
\usepackage{verbatim}
\usepackage{graphicx}
\usepackage{cite}
\usepackage{algorithm}
\usepackage{booktabs}
\usepackage[noend]{algpseudocode}
\usepackage{subcaption}

\begin{document}

\title{DeepAir: A Multi-Agent Deep Reinforcement Learning Based Scheme for an Unknown User Location Problem}

\author{Baris Yamansavascilar, Atay Ozgovde, and Cem Ersoy}

\maketitle

\begin{abstract}

The deployment of unmanned aerial vehicles (UAVs) in many different settings has provided various solutions and strategies for networking paradigms. Therefore, it reduces the complexity of the developments for the existing problems, which otherwise require more sophisticated approaches. One of those existing problems is the unknown user locations in an infrastructure-less environment in which users cannot connect to any communication device or computation-providing server, which is essential to task offloading in order to achieve the required quality of service (QoS). Therefore, in this study, we investigate this problem thoroughly and propose a novel deep reinforcement learning (DRL) based scheme, DeepAir. DeepAir considers all of the necessary steps including sensing, localization, resource allocation, and multi-access edge computing (MEC) to achieve QoS requirements for the offloaded tasks without violating the maximum tolerable delay. To this end, we use two types of UAVs including detector UAVs, and serving UAVs. We utilize detector UAVs as DRL agents which ensure sensing, localization, and resource allocation. On the other hand, we utilize serving UAVs to provide MEC features. Our experiments show that DeepAir provides a high task success rate by deploying fewer detector UAVs in the environment, which includes different numbers of users and user attraction points, compared to benchmark methods. 


\end{abstract}


\section{Introduction}

The widespread utilization of cloud computing after nearly two decades has brought about many opportunities for both companies and end-users that benefit from task offloading, content caching, and resource allocation. Especially throughout its computational advantages, cloud computing has provided computing capacity, reliability, and robustness for the offloaded tasks which otherwise may not be solved on the user devices \cite{yamansavascilar2024air, akhlaqi2023task}. However, even though it provides important opportunities, many other computing paradigms including Cloudlet \cite{satyanarayanan2009case}, Edge Computing \cite{chen2015efficient, baktir2017can}, and Fog Computing \cite{computing2016fog} emerged in the last decade since cloud computing cannot meet the low latency requirements of novel user applications due to the wide area network delay (WAN) \cite{laroui2021edge}.

Among those emerged edge solutions, multi-access edge computing (MEC) \cite{wang2018power} has become an extensively used technology since it provides low latency and computation power for intensive tasks. Therefore, it is deployed for different application types including healthcare, video analytics, smart home, and virtual reality \cite{mach2017mobile, mao2017survey}. Nevertheless, the fixed infrastructure of MEC prevents its utilization for dynamic scenarios in which the number of users/requests increases suddenly because of an event or a disaster. For example, a sport or concert event, in which there are thousands of new users, can exceed the capacity of the existing MEC infrastructure and therefore the quality of service (QoS) of users can be heavily affected as the corresponding tasks cannot be executed properly.

\begin{figure}[t]
\centering
\frame{\includegraphics[scale=0.045]{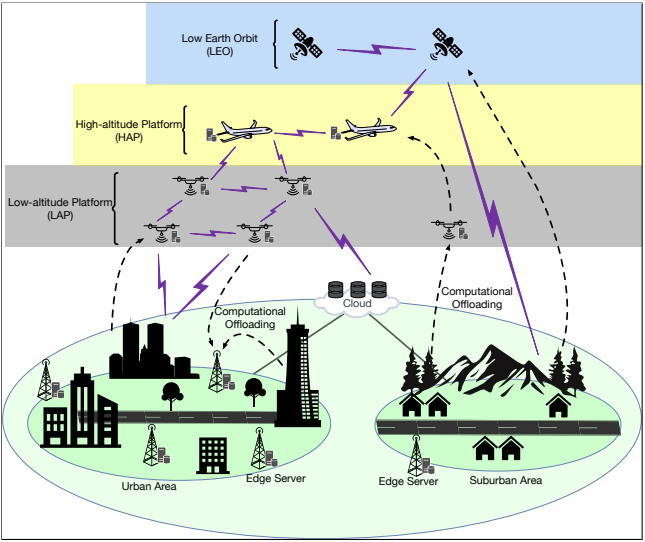}}
\caption{An air computing environment with different air platforms.}
\label{AirComputing}
\end{figure}

To meet the dynamic capacity requirements, UAVs have recently been deployed  along with other air vehicles in different air platforms under varied names such as aerial radio access network (ARAN), and air computing \cite{yamansavascilar2024air, dao2021survey}. An air computing environment is depicted in Figure \ref{AirComputing}. The communication between those different air platforms throughout the corresponding air vehicles brings about new research opportunities to meet those requirements considering the application QoS and user Quality of Experience (QoE) \cite{cao2018airborne}. Thus, different application profiles can benefit from various advantages of this new 3D dynamic computing paradigm.

Among those different air vehicles, UAVs are the most studied units since their deployment is easier considering their energy consumption, flying altitude, and configuration \cite{li2018uav}. To this end, they are used for dynamic capacity enhancement in environments in which the fixed capacity would not be sufficient to meet the application requirements of an increasing number of users. Therefore, this feature solves variety of problems such as communication in a disaster site, and enhancing services in infrastructure-less environments \cite{guo2021service, liu2018space}. Moreover, their deployment provides significant vertical networking opportunities such as high mobility support, coverage, latency, and two-way task offloading \cite{cheng2018air}. As a result, the requirements of users living in urban, suburban, and rural areas can be met efficiently through these vertical networking opportunities.

There are many studies in which UAVs are used as flying computational units to assist either deployed edge servers or are solely deployed to enhance network capacity for task offloading \cite{liu2018space, baltaci2021survey}. Since the battery and CPU capacity of the end users would not be sufficient to process the corresponding tasks, UAVs therefore can be used as a flying edge server. However, since there are many different scenarios, various methods and algorithms are developed to meet the service requirements. Deep Reinforcement Learning (DRL) is one of those methods that is applied in the literature since the traditional heuristic methods and convex optimization cannot solve the corresponding dynamic problems \cite{xiao2021leveraging}. To this end, DRL solutions would be used for trajectory optimization, energy-efficient offloading, UAV placement, and generic task offloading. 

\subsection{Motivations}

User connectivity is a primary issue in accessing the related resources for task offloading and service differentiation. However, in order to provide a required service, the corresponding technology such as edge or UAV should first detect the user, and then the connection should be established. However, in an environment which is infrastructure-less and user locations are unknown, providing those services is a crucial technical challange.

Even though UAVs are used in many different cases, their utilization in an infrastructure-less environment in which users cannot connect to any cellular operator, edge/cloud server, or satellite has not been investigated properly. That environment can be a disaster site, wilderness, or a natural area that is open to visitors. In such an environment, the detection of user locations, localization, and then the measurement of required capacity for user tasks are major issues to ensure the necessary service. Therefore, in this study, we focus on an environment in which there are users at unknown locations where we locate them through a novel method using DRL. Afterwards, as the second step, we estimate the corresponding requirements of the connected users at detected locations and decide how many UAVs are needed in those corresponding areas.

On the other hand, as detailed by Bai et al. \cite{bai2023towards}, DRL-based UAV studies have four main categories, and most of the studies in the literature focus on a subset of those categories. Therefore, in addition to the challenging environment, providing a solution for each of those categories is our overall goal.

\subsection{Contributions}

In this study, we develop a DRL-based scheme, DeepAir, which takes unconnected users' emitted RSSI signals into account as a reward and then finds the corresponding user attraction points in the environment. Since the number of user attraction points could not be known, DeepAir iteratively detects those points. Afterwards, based on the quality of those detected locations, users that are in the coverage can connect to agents, which we also call detector UAVs. After these phases, the corresponding user task profiles are extracted through the requests of connected users. Next, the required capacity based on those task profiles is computed, and then the necessary number of serving UAVs, which have MEC features, is measured for the detected user-concentrated areas. Thus, we increase the task success rate based on their Service Level Agreement (SLA) requirements. The main contributions of our paper are as follows. 

\begin{itemize}

\item We investigate the case of providing service to end-users in an infrastructure-less environment considering their QoS requirements which is not thoroughly investigated before. In this environment, users cannot connect to any operator or server, therefore their locations are not known.

\item In DRL-based UAV studies, a subset of four categories, including (1) sensing, (2) localization, (3) resource allocation, and (4) UAV-assisted MEC, are mainly considered in most of the cases \cite{bai2023towards}. In this study, we take all of these categories into account using our novel approach. 

\item In order to provide such a system, we develop a DRL-based multi-agent scheme, DeepAir, which is a novel approach that iteratively detects user locations in the environment using unconnected users' additive RSSI as the reward. DeepAir performs this operation iteratively, sending a single agent (detector UAV) to the environment for each iteration, since the number of user concentrated areas cannot be known.



\end{itemize}

The rest of this paper is organized as follows. In Section II, we elaborate on the related works including task offloading, UAVs, and DRL. We provide the system model and problem formulation in Section III. In Section IV, we introduce DeepAir providing technical and theoretic discussions. We show the experimental results in Section V. We discussed our observations through experiments in Section VI. Finally, we conclude our study in Section VII. We list the abbreviations used throughout the paper in Table \ref{listAbbreviations}.

\begin{table}[t]
\caption{List of abbreviations}
\label{listAbbreviations}
\centering
\begin{tabular}{| l | p{6cm} |}
\hline
\textbf{Notation} & \textbf{Description}\\
\hline
CF & Community Flying \\
DQN &Deep Q-Network \\
DRL&Deep Reinforcement Learning\\
MDP&Markov Decision Process\\
MEC & Multi-Access Edge Computing \\
QoS& Quality of Service\\
QoE& Quality of Experience\\
UAV & Unmanned Aerial Vehicles\\
SLA& Service Level Agreement\\
WAN & Wide Area Network\\
\hline
\end{tabular}
\end{table}

\section{Related Works}

\begin{table*}[t]
\caption{Comparison Between Related Studies}
\label{comparedStudies}
\centering
\begin{tabular}{ c | c  | c | c |  c | c | c |  c  }

\cline{2-7}
 \hline
\textbf{Study} & \textbf{Multi-UAV} & \textbf{\shortstack{Unknown \\ User Location}} & \textbf{Sensing}  & \textbf{\shortstack{Resource \\ Allocation} } & \textbf{Localization}  &  \textbf{\shortstack{UAV-assisted \\ MEC}  } \\
\hline
\cite{wang2020multi} &  \checkmark &     &   &   &  \checkmark & \checkmark \\
\hline
\cite{chang2022trajectory} &  \checkmark &    &  & \checkmark &  \checkmark & \\
\hline
\cite{zhang2023multi} &  \checkmark &   \checkmark & \checkmark &  & \checkmark   &\\
\hline
\cite{zhang2024enhancing}&  \checkmark & \checkmark  &  &  &\checkmark  &\\
\hline
\cite{cheng2024trace}&  \checkmark & \checkmark  & \checkmark  &  &\checkmark  &\\
\hline
\cite{ning2023multi}&  \checkmark &   &  &  &\checkmark  &\\
\hline
\cite{hao2024joint}&  \checkmark &   &  & \checkmark &\checkmark  & \checkmark\\
\hline
\cite{shi2024task}&  \checkmark &   &  &  &\checkmark  & \checkmark \\
\hline
\cite{zhang2023deep}&   &   &  & \checkmark &\checkmark  & \checkmark \\
\hline
\cite{oubbati2021dispatch}& \checkmark  &   &  & \checkmark &\checkmark  &  \\
\hline
Our study& \checkmark &  \checkmark  & \checkmark  & \checkmark   & \checkmark   & \checkmark \\
\hline
 \end{tabular}
\end{table*}

We surveyed the related research papers considering our DeepAir implementation and scenario in which user locations are not known and UAVs are primarily used for sensing, and localization. We conducted our research considering various high-impact journals and conferences.

In \cite{wang2020multi}, Wang et al. focused on the fairness-related optimization of user equipments considering geographical fairness, load, and overall energy consumption. To perform this, they developed a multi-agent deep reinforcement learning based trajectory control algorithm based on the Multi-Agent Deep Deterministic Policy Gradient (MADDPG) algorithm. They used average the fairness index and overall energy consumption as their success metric. Chang et al. proposed a trajectory design and resource allocation scheme based on multi-UAVs \cite{chang2022trajectory}. To this end, they considered the joint user association, power allocation, and trajectory design to maximize the system utility. They used a multi-agent DRL method which performs centralized learning and decentralized execution. In \cite{zhang2023multi}, Zhang et al. investigated user localization through UAV swarms in the case of a disaster-affected ground that has no base station. Their goals were to increase the efficiency of the task and to minimize the energy consumption of the UAVs. They proposed a multi-Agent DRL approach whose initial route is based on the probability distribution map of the users.

In \cite{zhang2024enhancing}, Zhang et al. investigated a multi-UAV cooperative reconnaissance and search (MCRS) scheme for the localization of static targets. Therefore, they designed a belief probability map model based on Dempster-Shafer (DS) evidence theory and then proposed a new DRL algorithm called Double Critic Deep Deterministic Policy Gradient (DCDDPG). DCDDPG takes the belief probability map into account and uses the MADDPG approach by utilizing two critic networks. They evaluated their system performance based on decreasing uncertainty, and increasing number of targets found. Chent et al. focused on energy-efficient and dynamic multi-UAV coverage control for disaster areas \cite{cheng2024trace}. They developed a trace pheromone-based mechanism through the MADDPG algorithm in order to provide non-overlapping coverage. Based on reduced overlapping UAVs, they could achieve energy efficiency. They evaluated the performance of their system using average coverage rate, normalized average energy consumption, and coverage efficiency. In \cite{ning2023multi}, Ning et al. proposed a UAV trajectory optimization scheme considering user-differentiated services. Their goal was to minimize both the computational costs of users and UAVs. They solved the problems, including unknown information about the corresponding services, and UAV trajectory, by using a multi-agent DRL approach. In the end, they measured the overall cost of UAVs and users, respectively, as the performance metric.

In \cite{hao2024joint}, Hao et al. developed a DRL-based multi-UAV solution for task offloading problem. Therefore, they investigated a UAV-assisted MEC system considering energy consumption and task delay. Shi et al. focused on the optimization of the UAV trajectories and offloading strategies of users jointly \cite{shi2024task}. Hence, they investigated multi-UAV collaboration considering UAV-assisted MEC. To this end, they used a multi-agent DRL approach since the corresponding joint optimization requires non-convex operation. In \cite{zhang2023deep}, Zhang et al. investigated UAV-assisted communications using a single UAV and multiple users. To this end, they proposed a proximal policy optimization (PPO) based DRL algorithm in order to adjust the direction, altitude, and speed of the UAV. In addition to these adjustments, they took the QoS requirements of the users into account regarding service time minimization. Oubbati et al. proposed a DRL-based multi-UAV optimization method called DISCOUNT in order to utilize UAVs as relays in vehicular ad hoc networks (VANETs) \cite{oubbati2021dispatch}. Accordingly, they considered energy consumption, coverage, and routing performance to cover sparse areas in the network.

To the best of our knowledge, the unknown user location case is not deeply investigated by the related studies that utilize UAVs to provide required services in an environment. Thus, we provide a novel DRL-based approach, DeepAir, which locates the users through their RSSI using UAVs iteratively. Afterwards, we compute the QoS requirements of connected users in the detected areas and provide the corresponding serving UAVs. On the other hand, our novel scheme provides sensing, resource allocation, localization, and UAV-assisted MEC features that the related studies ensure only a subset of them. Our main differences between the related studies are given in Table \ref{comparedStudies}.

\section{System Model and Problem Formulation}
\label{SysModel}








\begin{table}[t]
\caption{List of main notations}
\label{NotationList}
\centering
\begin{tabular}{| l | p{6cm} |}
\hline
\textbf{Symbol} & \textbf{Definition}\\
\hline
\hline
$M$ &Number of users\\
\hline
$N_s$ &Number of serving UAVs\\
\hline
$N_d$ &Number of detector UAVs\\
\hline
$r_{n_d}$ &Horizontal radius of a detector UAV\\
\hline
$r_{n_s}$ &Horizontal radius of a serving UAV\\
\hline
$W_m$ &Task of user $m$\\
\hline
$D_m$ & Size of task $W_m$\\
\hline
$C_m$ & Required CPU cycle for task $W_m$\\
\hline
$\lambda_m$ & Arrival rate of task $W_m$\\
\hline
$T_{m}^{max}$ & Maximum tolerable delay of task $W_m$\\
\hline
$d_n(t)$ & Flight distance of UAV $n$ at time $t$\\
\hline
$\vartheta_n(t)$ & Flight angle of UAV $n$ at time $t$\\
\hline
$f_m$ & Computational capacity of user $m$\\
\hline
$f_{n_s}$ & Computational capacity of serving UAV $n_s$\\
\hline
$T_m^{loc}$ & Computation delay of user $m$\\
\hline
$T_{n_s}^{UAV}$ & Computation delay of serving UAV $n_s$\\
\hline
$T_t^{UAV}$ & Total delay including queueing and computation at serving UAV $n_s$\\
\hline
$T_q$ & Queueing delay at serving UAV $n_s$\\
\hline
$T_{m, n_s}$ & Transmission delay between user $m$ and serving UAV $n_s$\\
\hline
$h_{m, n_d} (t)$ & Channel gain between user $m$ and detector UAV $n_s$ after path loss\\
\hline
$g_{m, n_d}$ & Channel power gain between user $m$ and detector UAV $n_s$\\
\hline
$d_{m, n_d}$ & Distance between user $m$ and detector UAV $n_d$\\
\hline
$v_{m, n_s}$ & Data rate (bit/sec) between user $m$ and serving UAV $n_s$\\
\hline
$H(t)$ & Cumulative signal strength on detector UAV $n_s$ at time $t$\\
\hline
$t_{w_m}$ & Total delay for task $W_m$\\
\hline
$\beta{m, n_s}$ & Offloading decision variable between user $m$ and serving UAV $n_s$\\
\hline
$\alpha_{w}$ & Task completion variable\\
\hline
\end{tabular}
\end{table}

We consider an environment as a set of users denoted as $\mathcal{M} =  \{1, 2, ..., M\}$, a set of serving UAVs represented as $\mathcal{N}_s = \{1, 2, ..., N_s\} $, and a set of detector UAVs specified as $\mathcal{N}_d = \{1, 2, ..., N_d\} $.  Detector UAVs are used for sensing and localization of users in the environment whose locations are unknown. Serving UAVs, on the other hand, are used as a computational resource for offloaded tasks of users. In other words, serving UAVs are flying edge servers in the environment.  
Each UAV type has a horizontal radius, $r$, for communicational or computational coverage. Each user $m \in \mathcal{M}$ randomly produces a computation-intensive task $W_m = (D_m, C_m, \lambda_m, T_{m}^{max})$, where $D_m$ is the size of the task as bits, $C_m$ is the required CPU cycle for processing as cycle/bit, $\lambda_m$ is the arrival rate as task/sec, and $T_{m}^{max}$ is the maximum tolerable latency as seconds. For convenience to read, the list of main notations used in formulations is given in Table \ref{NotationList}. 

In the environment, the location of a UAV $ n \in \mathcal{N}_s$ or $n \in  \mathcal{N}_d$ can be denoted as $u_n(t) = [ x_n(t), y_n(t), z_n(t) ]$, where  $x_n(t), y_n(t),$ and  $z_n(t)$ are the $X, Y, Z$ coordinates at time time $t$. Therefore, the position of UAV $n$ at the next time step for a horizontal flight can be expressed as 

\begin{equation}
\label{UAV-Coordinates-x}
	x_n(t+1) = x_n(t) + d_n(t) \times cos(\vartheta_n(t))
\end{equation}

\begin{equation}
\label{UAV-Coordinates-y}
	y_n(t+1) = y_n(t) + d_n(t) \times sin(\vartheta_n(t))
\end{equation}
where $d_n(t)$ is the flight distance, and $\vartheta_n(t) \in [0, 2\pi]$ is the flight angle. Moreover, the following movement constraints should be satisfied during the horizontal flight considering the area of the environment

\begin{equation}
\label{UAV-environment-constraint-1}
	0 \le x_n(t) \le X_{max}
\end{equation}

\begin{equation}
\label{UAV-environment-constraint-2}
	0 \le y_n(t) \le Y_{max}
\end{equation}
where $X_{max}$ and $Y_{max}$ are the maximum lengths of the environment. Similarly, to avoid collision between any two UAVs, including serving and detector UAVs, minimum distance should be satisfied as follows
\begin{equation}
\label{UAV-minDistance}
|| u_i(t) - u_j(t) || \ge d_{min} \quad \forall i, j, i \neq j
\end{equation}
 where $d_{min}$ denotes the minimum distance. On the other hand, location of a user $m \in \mathcal{M} $ at time $t$ is represented as $L_m(t) = [x_m(t), y_m(t), 0]$ where  $x_n(t),$ and  $y_n(t)$ are the $X$, and $Y$ coordinates, respectively. Since users are in the ground, the Z coordinates are zero. 


If a user $ m \in \mathcal{M}$ processes its task locally, the corresponding local computation delay is measured as follows

\begin{equation}
\label{local-execution}
	T_m^{loc} = \frac{D_m \times C_m}{f_m}
\end{equation}
where $f_m$ is the computational capacity of user $m$ as CPU cycle per second. On the other hand, if the task $W_m$ is offloaded to a serving UAV $n_s \in \mathcal{N}_s$, the computation delay at $n_s$ is measured as 

\begin{equation}
\label{UAV-execution}
	T_{n_s}^{UAV} = \frac{D_m \times C_m}{f_{n_s}}
\end{equation}
where $f_{n_s}$ is the computational capacity of UAV $n_s$ as CPU cycles per second. Since multiple users can be connected and therefore offload their tasks to a serving UAV, $M/M/1$ queueing model is used for the overall delay measurement at the corresponding serving UAV as  

\begin{equation}
\label{UAV-queueing}
	T_{t} = \frac{\sum_{m}^{\mathcal{M}_{sub}} D_m \times C_m}{f_{n_s} - \sum_{m}^{\mathcal{M}_{sub}}  (\lambda_m \times D_m \times C_m)}
\end{equation}
where $\mathcal{M}_{sub}$ denotes a subset of users concentrated in the corresponding area. Therefore, queueing delay at the serving UAV is measured as

\begin{equation}
\label{UAV-queueing-delay}
	T_{q} = T_{t} - T_{n_s}^{UAV}
\end{equation}

Considering the task offloading case for task $W_m$, transmission delay between a user $m \in \mathcal{M}$ and a serving UAV $n_s \in \mathcal{N}_s$ is measured as 

\begin{equation}
\label{transmissionDelay}
	T_{m, n_s} = \frac{D_m}{v_{m, n_s}}
\end{equation}
where $v_{m, n_s}$ is the data rate between $m$ and $n_s$ as bit/sec. It is important to note that users connect and communicate multiple serving UAVs via orthogonal frequency-division multiple access (OFDMA). Therefore, the transmission interference between different users can be ignored.

Channel gain, which indicates to the measurement of the strength of the signal between the transmitter and receiver in wireless communication, is computed between a user $m$ and a detector UAV $n_d \in \mathcal{N}_d$ using free-space path loss model as 

\begin{equation}
\label{ChannelGain}
	h_{m, n_d} (t)= \frac{g_{m, n_d}}{| d_{m, n_d}(t) | ^2}
\end{equation}

where $g_{m, n_d}$ denotes the channel power gain between the user $m$ and detector UAV $n_d$, and $d_{m, n_d}(t)$ is the distance at time $t$. Considering the cumulative signal strength of users at a detector UAV $n_d$, it is measured as 

\begin{equation}
\label{CumulativeChannelGain}
	H (t)= \sum_{m}^{\mathcal{M}} h_{m, n_d} (t)
\end{equation}

Since serving UAVs are sent to the locations that detector UAVs has already provided, we assume that the channel quality between users and serving UAVs is already above an acceptable threshold and therefore is not included in the formulation.

\subsection{Problem Formulation}

In the environment, the total delay for a task of user $m$ is computed as 

 \begin{equation}
\label{overallDelay}
	t_{w_m} = t_{network} + t_{service}
\end{equation}
where $t_{network}$ is the network delay, and $t_{service}$ is the service delay. The network delay is computed as  

 \begin{equation}
\label{networkDelay}
	t_{network} = 
	\begin{cases}
     T_{m, n_s} , & \text{if} \quad \beta_{m, n_s} = 1 \\
     0, & \text{if} \quad \beta_{m, n_s} = 0
    \end{cases}
\end{equation}
where $\beta_{m, n_s}$ is a binary variable that indicates whether the task is offloaded to a serving UAV or not. 
While calculating the network delay, the propagation delay is ignored. The service delay on the other hand is computed as 

 \begin{equation}
\label{serviceDelay}
	t_{service} = 
	\begin{cases}
     T_{n_s}^{UAV} + T_q , & \text{if} \quad \beta_{m, n_s} = 1 \\
     T_{n_s}^{loc}, & \text{if} \quad \beta_{m, n_s} = 0
    \end{cases}
\end{equation}

In our environment, a task of user $m$, $W_m$, is successfully completed if the total delay is lower than or equal to the maximum tolerable delay of the task. Hence, we define $\alpha_w$ as a success variable for task completion as 

 \begin{equation}
\label{taskSuccess}
	\alpha_w = 
	\begin{cases}
     1 , & \text{if} \quad t_{w_m} \le T_m^{max} \\
     0, & otherwise
    \end{cases}
\end{equation}

Thus, in this study, our main goal is to maximize the overall task success in the environment. To this end, our objective function is defined as 

\begin{equation}
	\label{objectiveFunction}
	\max  \sum_{m}^{\mathcal{M}}\sum_{w_m}^{W}\alpha_{w_m}
\end{equation}
subject to 

\begin{equation}
	\label{C1}
	\sum_{n_s}^{\mathcal{N}_s}\beta_{m, n_s} \le 1\quad \forall m \in \mathcal{M}
	\tag{Constraint 1}
\end{equation}

\begin{equation}
	\label{C2}
	d_{m, n_s} \le r_{n_s} \quad \forall m \in \mathcal{M}, \quad \forall n_s \in \mathcal{N}_s
	\tag{Constraint 2}
\end{equation}

\begin{equation}
	\label{C3}
	f_{n_s} - (\lambda_m \times D_m \times C_m) > 0  \quad \forall m, n_s
	\tag{Constraint 3}
\end{equation}
\begin{equation}
	\label{Ct}
	 \text{Equations } (\ref{UAV-environment-constraint-1}) - (\ref{UAV-minDistance})
	\tag{Constraint 4}
\end{equation}
where $d_{m, n_s}$ is the distance between user $m$ and serving UAV $n_s$. Constraint 1 represents that a task can be offloaded to only a single serving UAV even though the user can connect to multiple serving UAVs. Constraint 2 denotes that a user should be in the coverage of a serving UAV in order to connect it and then offload a task to it. Constraint 3 ensures that offloaded tasks cannot exceed the capacity of a serving UAV. Finally, Constraint 4 describes the movement constraints.



\section{DeepAir}


\begin{figure*}[!t]
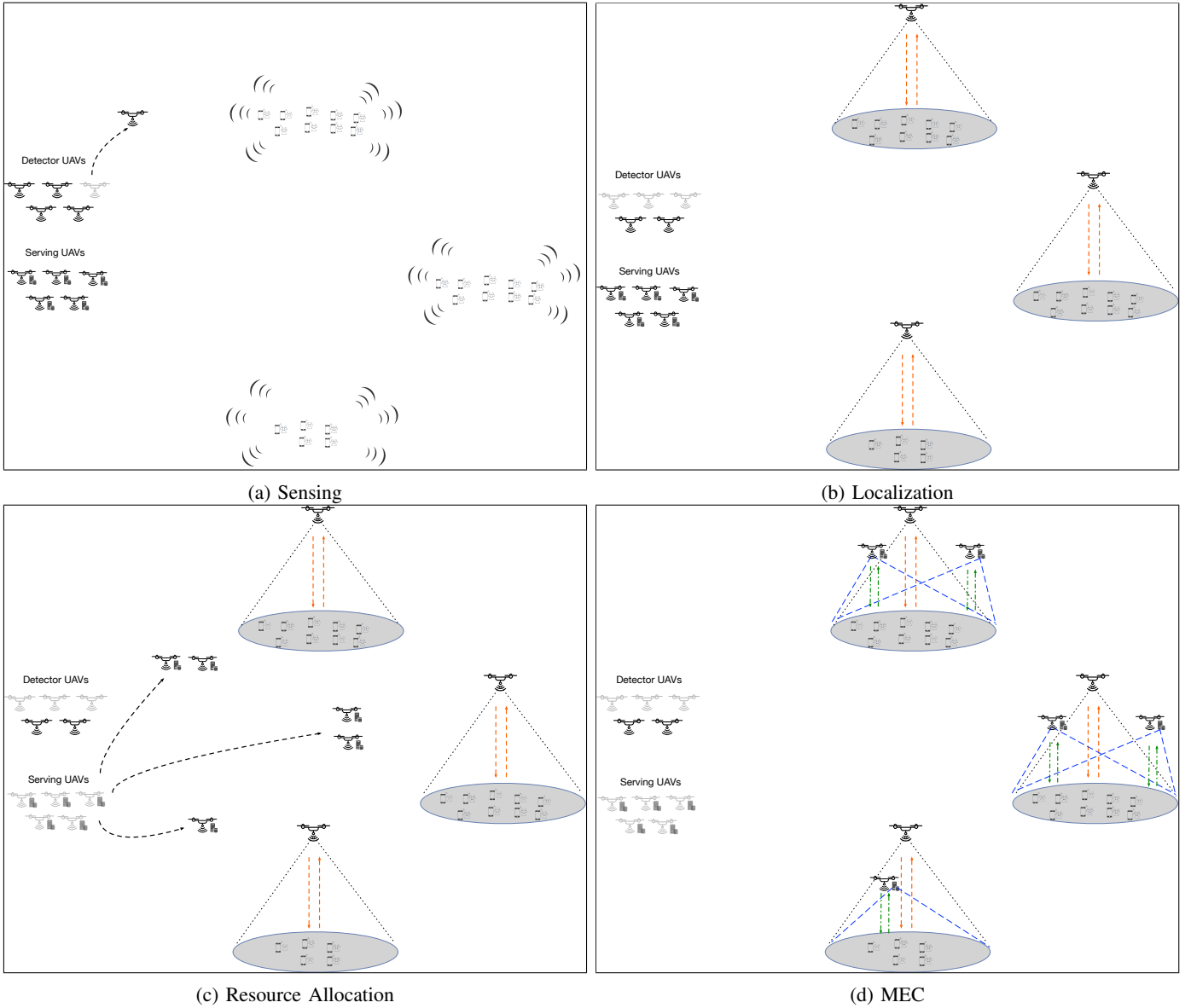

\begin{subfigure}{0.5\textwidth}
\frame{\includegraphics[width=\linewidth]{images/Sensing.pdf}}
\caption{Sensing} \label{sensing}
\end{subfigure}
\hspace*{\fill} 
\begin{subfigure}{0.5\textwidth}
\frame{\includegraphics[width=\linewidth]{images/Localization.pdf}}
\caption{Localization} \label{localization}
\end{subfigure}
\begin{subfigure}{0.5\textwidth}
\frame{\includegraphics[width=\linewidth]{images/ResourceAllocation.pdf}}
\caption{Resource Allocation} \label{ResourceAllocation}
\end{subfigure}
\hspace*{\fill} 
\begin{subfigure}{0.5\textwidth}
\frame{\includegraphics[width=\linewidth]{images/MEC.pdf}}
\caption{MEC} \label{MEC}
\end{subfigure}
\caption{Phases of DeepAir}
\label{DeepAir-Phases}
\end{figure*}

Our DeepAir operation has four main phases that should be considered to provide the required QoS for user tasks. As formulated in Section \ref{SysModel}, a user task is successfully completed if the total delay in the system is smaller than or equal to its maximum tolerable delay. Therefore, each phase, including sensing, localization, resource allocation, and MEC, is significant for the efficient operation in the environment. To this end, we use detector UAVs for the sensing, localization, and resource allocation phases. Note that a detector UAV is also used as a DRL agent in the environment to find the user-concentrated areas. On the other hand, based on the reporting of detector UAVs, we deploy serving UAVs for the offloading and processing of tasks as part of the MEC resources. The whole operation including four phases is depicted in Figure \ref{DeepAir-Phases}.

Each type of UAV in the environment can communicate with each other via a separate channel. Thus, they know their existing location. Moreover, they can also communicate with the base, which is at $(0, 0)$ coordinates, so that they can notify the existing situation in the corresponding areas in their horizontal coverage. As a result, the system can react to the events in those corresponding areas in real-time.

\subsection{Sensing}

\begin{figure}[!t]
\begin{subfigure}{0.42\textwidth}
\frame{\includegraphics[width=\linewidth]{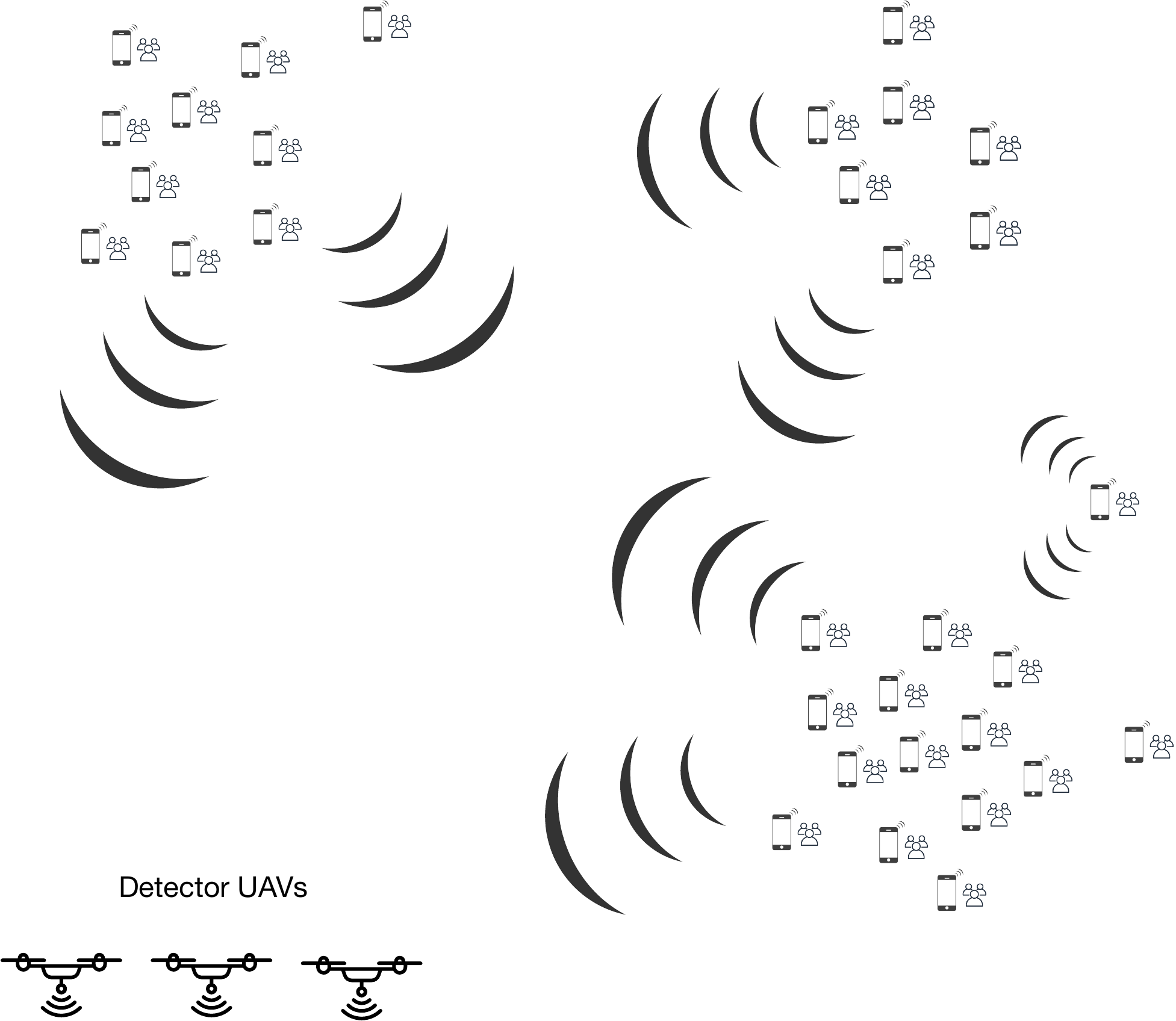}}
\caption{Environment consisting of users that emits connection signals} \label{emitted-1}
\end{subfigure}
\begin{subfigure}{0.42\textwidth}
\frame{\includegraphics[width=\linewidth]{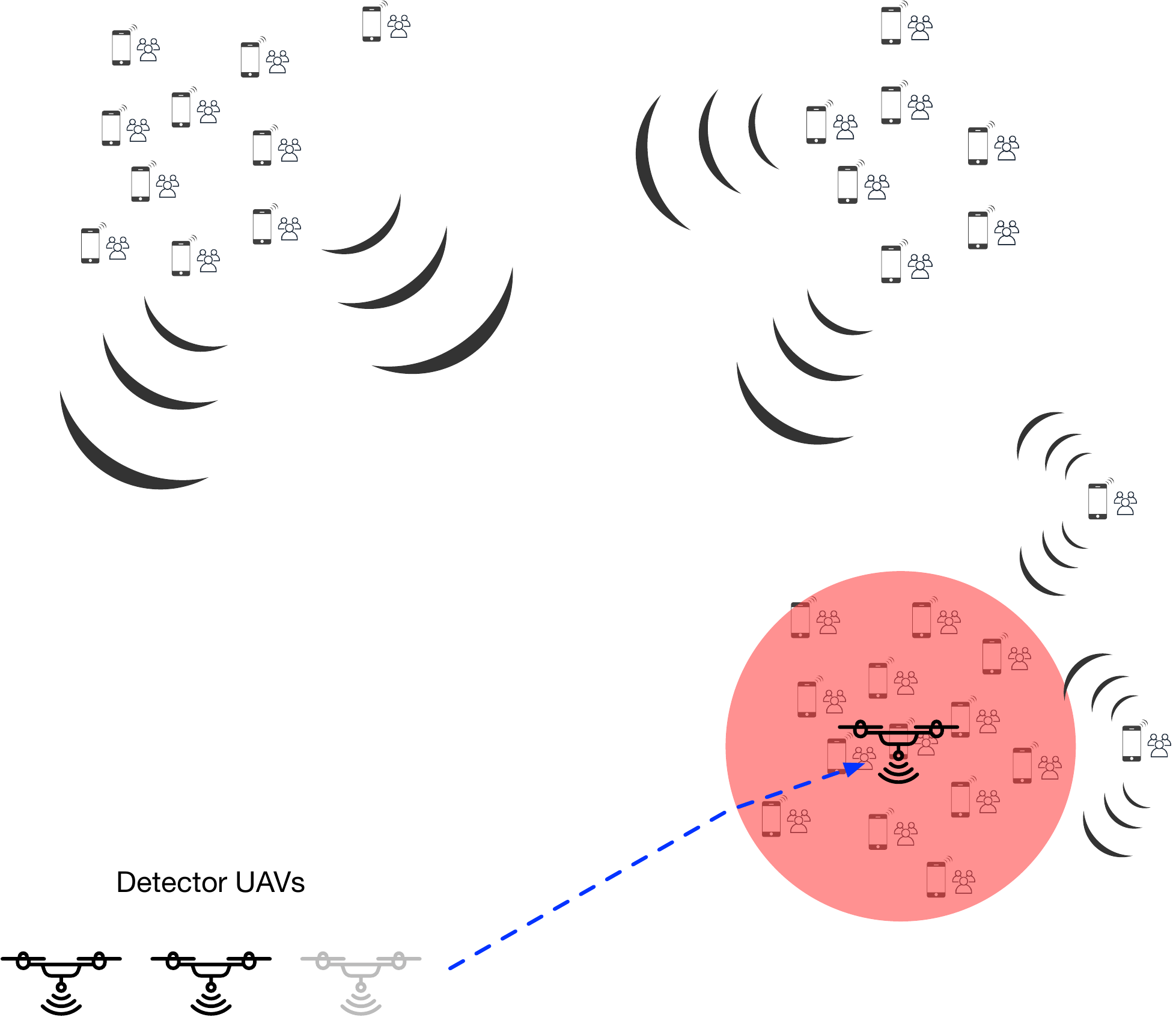}}
\caption{Emitted signals of connected users stop after the connection} \label{emitted-2}
\end{subfigure}
\caption{Depiction of the state of environment regarding the emitted signals}
\label{emitted}
\end{figure}

Due to the nature of the infrastructure-less deployment, initially, users in the environment are not connected to any system component, emitting only signals for a possible connection. Therefore, as shown in Figure \ref{emitted-1}, a detector UAV can sense signal strength at some point in the environment at time $t$. Based on the location of users and the detector UAV, that signal strength can change in different areas of the environment considering the cumulative signal strength measurements in Equations \ref{ChannelGain} and \ref{CumulativeChannelGain}.

In this study, we assume that there are different user attraction points in the environment whose locations are also unknown. Based on those user attraction points as shown in Figure \ref{emitted-1}, users are gathered in certain areas. Therefore, the corresponding additive signal strength would be higher when the detector UAV is close to that area. However, considering the fact that there would be many user attraction points whose locations could be in different parts of the environment, and some of those parts may have similar user densities, sensing levels can turn out to have identical or similar values for different points in the environment. Thus, this fact should be taken into account in the localization phase in which we use DRL.

\subsection{Localization}

In the localization phase, information gathered in sensing is initially used for the movement of detector UAVs (agents) and then to locate the corresponding user attraction points. One of the crucial factors here is that the number of user attraction points in the environment is also unknown along with the number of users, and user locations. Therefore, we cannot apply multiple agents simultaneously in the environment since if the number of deployed agents is less than the number of user attraction points then users at some of the attraction points cannot be detected and served. Thus, we apply an iterative approach using DRL as shown in Algorithm \ref{Phase-1}.

In Algorithm \ref{Phase-1}, we find the user attraction points in the environment. To perform this, we initially set a threshold for new connected users in an iteration. For each iteration we send a DRL agent to the environment flying from the base at $(0, 0)$ coordinates, and after the convergence it returns the corresponding location information along with how many new users are connected to it. If the number of connections is smaller than the threshold, the execution of the algorithm is terminated. Otherwise, we continue to send an agent to the environment. Note that when a user device connects to a detector UAV, it stops emitting the connection signal. As a result, the additive signal power would be less in a particular place in the next iteration for the agent. This situation is depicted in Figure \ref{emitted}.







\subsubsection{MDP Definition}

The DRL is based on MDP which is formally defined as a 4-tuple $<S, A, P, R>$ where

\begin{itemize}

\item $S$ is the set of states where $s \in S$

\item $A$ is the set of actions where $a \in A$

\item $P: S \times A \rightarrow P(S)$ is the state transition probability function that provides the probability of $P(s_t, s_{t+1}) = P(s_{t+1} = s' \mid s_t = s, a_t = a)$. This function denotes that the current state $s_t$ at time $t$ changes to the state $s_{t+1}$ by taking the action $a$.

\item $R: S \times A \times S \rightarrow R $ is the reward function. It defines the corresponding reward $\mathcal{R}$ at time $t$ by taking an action $a$ in a state $s_t$. Therefore, it can be also defined as $\mathcal{R}_t = R(s_t, a_t, s_{t+1})$.

\end{itemize}

Our environment provides the corresponding MDP definition through the state representation, action space, reward mechanism, and state transition. Therefore, an applied DRL algorithm in the environment using a detector UAV can create a policy $\pi$ based on a given state as  $\pi \left(a \mid s  \right) = P \left(a_t = a \mid s_t  = s \right)$. Through this policy, the agent can learn the dynamics in the environment for a given state and therefore takes the most effective action.

\begin{algorithm}[t]
    \caption{Finding Locations}\label{Phase-1}
    \begin{algorithmic}[1]
    	\State  \textbf{Input:} $Threshold$ and the environment
	\State  \textbf{Output:} Found Locations
        \State $isThresholdMet$ = True
        \State $locations$ = [  ]
        \While{$ isThresholdMet $}
        		\State $singleLocation$, $connectionCount$ = DDQN() 
		\If{$connectionCount$ is smaller than $Threshold$ }
			\State $isThresholdMet = False$
		\Else
			\State Add $singleLocation$ to $locations$
		\EndIf
	 \EndWhile
\Return $locations$

    \end{algorithmic}
\end{algorithm}

\subsubsection{State Space $S$}

Based on the Algorithm \ref{Phase-1}, there is only a single agent in the environment for each iteration. Therefore, the state at time $t$ consists of the current location of the agent as

\begin{equation}
	\label{StateDef}
	s(t) = u_n(t) = [ x_n(t), y_n(t), z_n(t) ]
\end{equation}

\subsubsection{Action Space $A$}

In our environment, the action space $A$ consists of five discrete actions considering the horizontal movements. Therefore, it is defined as

\begin{equation}
	\label{StateDef}
	a(t) = [ Left, Right, Up, Down, NoMove ]
\end{equation}

Based on this definition, the horizontal speed of each agent is fixed during their flight. Considering the selected action at time $t$, they can stay fixed at their horizontal coordinates by $NoMove$ action. Otherwise, they can move into four different areas of the environment.

\subsubsection{Reward Function $R$}

Based on the policy $\pi$, the agent takes the corresponding actions in the environment to maximize its cumulative reward. To this end, for a given state $s_t$, the agent maximizes the expected sum of future reward by applying policy $\pi(s_t)$ as follows

\begin{equation}
\label{ExpectedReward}
R_t =  \sum_{i=t}^{\infty} \gamma*R(s_i, s_{i+1}) 
\end{equation}
where $\gamma \in [0,1]$ is the discount factor that denotes the importance of the long-term rewards if its value is close to one. Otherwise, its value would be close to zero. Thus, the reward function is defined as follows based on the consideration above

 \begin{equation}
\label{RewardFunction}
	R(t) = 
	\begin{cases}
     H(t) , & \text{if} \quad \text{satisfying constraints}  \\
     -1, & \text{if} \quad otherwise
    \end{cases}
\end{equation}

\subsubsection{Application of DRL}

Since our action space is discrete, applying a value-based DRL algorithm is more convenient in our environment. Therefore, for each DRL agent, we implement Deep Q-Learning (DQN) algorithm \cite{arulkumaran2017deep}, which manifests a high success in many different environments with different state spaces.

In value-based DRL algorithms such as DQN, the agent should select the best state-action pair among different options through its policy by a given state $s_t$ by maximizing Q-function, $Q_\pi(s_t , a_t)$. Therefore, under the policy $\pi$, the Q-function is defined as

\begin{equation}
\label{Q-function}
Q_\pi(s_t , a_t) = \mathop{\mathbb{E}} \big[ R_t \mid s_i = s, a_i = a \big]
\end{equation}
which denotes the value of an action $a_t$ in a state $s_t$. Thus, an optimal policy is defined as selecting the highest valued action in each state

\begin{equation}
\label{OptimalPolicy}
\pi(s_t) = arg\max\limits_{a'}(Q(s_t, a'))
\end{equation}
where $a'$ indicates the set of all possible actions.
As a result, the value of a state-action pair is computed as

\begin{equation}
\label{DDQN-Value}
z_t = R_{a_t}(s_t, s_{t+1}) + \gamma * Q(s_{t+1}, arg\max\limits_{a'}Q(s_{t+1},a'; \theta_t);\theta_t)
\end{equation}
where $\theta_t$ represents the Q-network parameters. On the other hand, considering the convergence through the Q-network, the agent minimize the temporal difference error, $\delta_t$, of $Q_\pi(s_t, a_t)$ regarding $z_t$:

\begin{equation}
\label{targetError}
\delta_t = \mid Q(s_t, a_t) - z_t  \mid
\end{equation}

\subsection{Resource Allocation}


After the completion of sensing and localization phases through detector UAVs, the next step is the resource allocation for serving UAVs which provide computational serving capabilities. Since a serving UAV has a limited capacity, measuring how many of them should be deployed is the main problem in this phase. 

As users have already connected to the corresponding detector UAVs, and have started to send their task offloading requests, the system can create a task profile in those user-connected areas by conducting several capacity calculations. To this end, we perform a capacity calculation that includes the task profile of each user as we defined $W_m = (D_m, C_m, \lambda_m, T_{m}^{max})$. The measurements are essentially based on Equation \ref{UAV-queueing} as the delay at serving UAVs is based on $M/M/1$ queueing model.

After the measurement of the required number of serving UAVs for each detected area, the other important issue is the deployment of available serving UAVs into those areas, each of which may have different task profiles. Note that the available serving UAVs may not meet the total required serving UAVs in the environment. In this case, available serving UAVs are first deployed to the areas which require the lowest 
$T_{m}^{max}$. In other words, if an area has a higher need for serving UAVs, that area has a priority. On the other hand, if required numbers of serving UAVs are equal for the corresponding areas, then a round-robin approach is applied for the deployment. 

\subsection{MEC}

After the resource allocation, the next and final phase is providing the MEC features to users. To this end, users offload their tasks to serving UAVs and expect a valid service without violating the task's maximum tolerable delay, $T_{m}^{max}$. 

Note that a user in the corresponding area can connect to multiple serving UAVs simultaneously. To this end, we assume that a user is informed by those serving UAVs via separate channel about the current queueing condition. Hence, a user connected to multiple serving UAVs can select the serving UAV for task offloading regarding minimum delay.

\section{Performance Evaluation}

We conducted experiments using a discrete event simulation for the performance evaluation. In these experiments, we have an environment whose size is $500 \times 500$ $m^2$. In this environment, there are various number of user attraction points around which the user densities are higher. Note that the location of those users and attraction points are initially not known by the system. The corresponding simulation parameters are given in Table \ref{SimParameters}. Throughout the experiments, we used Python 3.10. Moreover, we used PyTorch version 2.2.0 for the training of DRL agents. 

\begin{table}[!t]
\caption{Simulation Parameters}
\label{SimParameters}
\centering
\begin{tabular}{|l|l|}
\hline

\textbf{Parameter} & \textbf{Value}\\
\hline
\hline
Number of User Attraction Points  & [3, 5] \\
\hline
Size of a task ($D_m$) & 500 Kb\\
\hline
Required CPU cycles of a task ($C_m$)& 90 cycles/bit\\
\hline
Arrival rate of a task ($\lambda_m$) & 0.30 task/sec\\ 
\hline
Maximum tolerable delay of a task ($T_{m}^{max}$) & 1 sec\\
\hline
Capacity of a serving UAV ($f_{n_s}$) & 300 cycles/sec\\
\hline
UAV Radius ($r_n$) & 75 m\\
\hline
Data Rate ($v_{m, n_s}$) & 100 Mbps\\
\hline
Channel Power Gain ($g_{m, n_d}$) & $1.42 \times 10^{-4}$\\
\hline
New Connected User Threshold & 1 \\
\hline
$X_{max}$ & 500 m \\
\hline
$Y_{max}$ & 500 m \\
\hline
Altitude of UAVs ($z_n(t)$)& 200 m \\
\hline
\end{tabular}
\end{table}

In our experiments, we assumed that each user in the environment produces a task with the parameters $D_m$, $C_m$, $\lambda_m$, and $T_m^{max}$. Moreover, each produced task should be offloaded to one of the serving UAVs since we assumed that the computational capabilities of user devices are not sufficient to meet the task requirements. Similarly, each detector and serving UAV is identical in terms of radius, and altitude. Considering the offloading, we ignored the propagation delay for simplicity. We repeated our experiments 50 times with different seeds. The duration of each experiment in simulator time was 1000 seconds. 


\subsection{Training Step}

Regarding Algorithm 1, we train a DRL agent through detector UAVs for each iteration as long as it provides new connections. To this end, we tested several hyperparameters throughout the training step in order to achieve convergence for different user distributions in the environment considering the final model. Generally, we used random search based on our experience in the domain \cite{bergstra2012random}.

\begin{table}[!t]
\caption{Hyperparameters of DRL Algorithm}
\label{Hyperparameters}
\centering
\begin{tabular}{|l|l|}
\hline
\textbf{Parameter} & \textbf{Value}\\
\hline
\hline
Learning Rate & 0.005\\
\hline
Discount Factor & 0.99\\
\hline
Initial Exploration & 1\\
\hline
Exploration Factor & 0.99\\
\hline
Final Exploration & 0.01\\
\hline
Replay Memory Size & 1000000\\
\hline
Batch Size & 64\\

\hline
\end{tabular}
\end{table}

The neural network in our final model consists of three layers each of which includes 128 neurons. Rectified Linear Unit (ReLU) is applied for each neuron as the activation function. We used mean squared error (MSE) for the loss, and stochastic gradient descent (SGD) as the optimizer. On the other hand, for the initial exploration, exploration factor, final exploration, and replay memory size, we used well-known values from the literature as given in Table \ref{Hyperparameters}.

\begin{figure}[!t]
\centering
\includegraphics[scale=0.6]{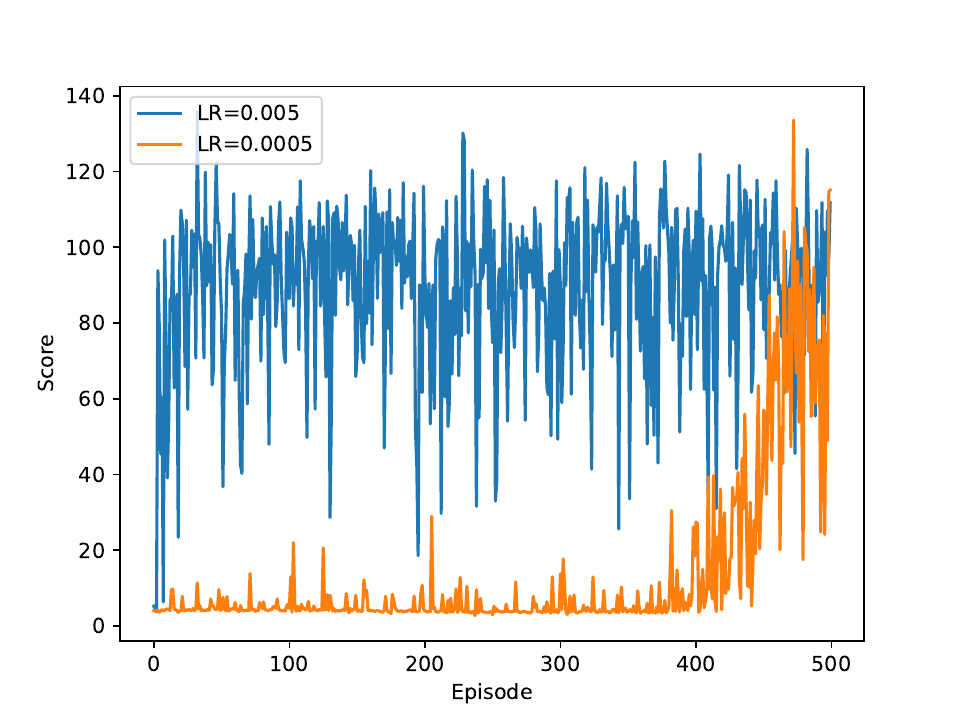}
\caption{Effect of learning rate on scores for each episode using 100 users.}
\label{episodes}
\end{figure}

Since the learning rate is the most crucial hyperparameter for the performance of the model, we exclusively tested different settings. Thus, we determined $0.005$ as the final value for the neural network of the agents. Figure \ref{episodes} shows the effect of two different learning rates over episodes. As shown in the figure, a lower value of learning rate, such as $0.0005$, causes a late convergence in the environment. On the other hand, if we use $0.005$ as the learning rate, the model converges earlier which provides time efficiency.

\subsection{Competitors}

We used two competitors as benchmark methods namely the Community Flying (CF) and Random methods similar to \cite{ning2023multi}. In CF, we divided the environment into equal communities, and the center of each community was evaluated as a possible user attraction point. Accordingly, detector UAVs are sent to those centers for possible connections and corresponding QoS measurements. 
Afterwards, the required number of serving UAVs is deployed based on the needs of those areas. On the other hand, in the random method, the possible attraction centers are selected randomly in the environment based on the available number of detector UAVs. We named the random methods based on the available number of detector UAVs as in the case of CF. To this end, for example, if we use 8 detector UAVs, then the method is named as Random-8.


\begin{figure}[t]
\centering
\includegraphics[scale=0.6]{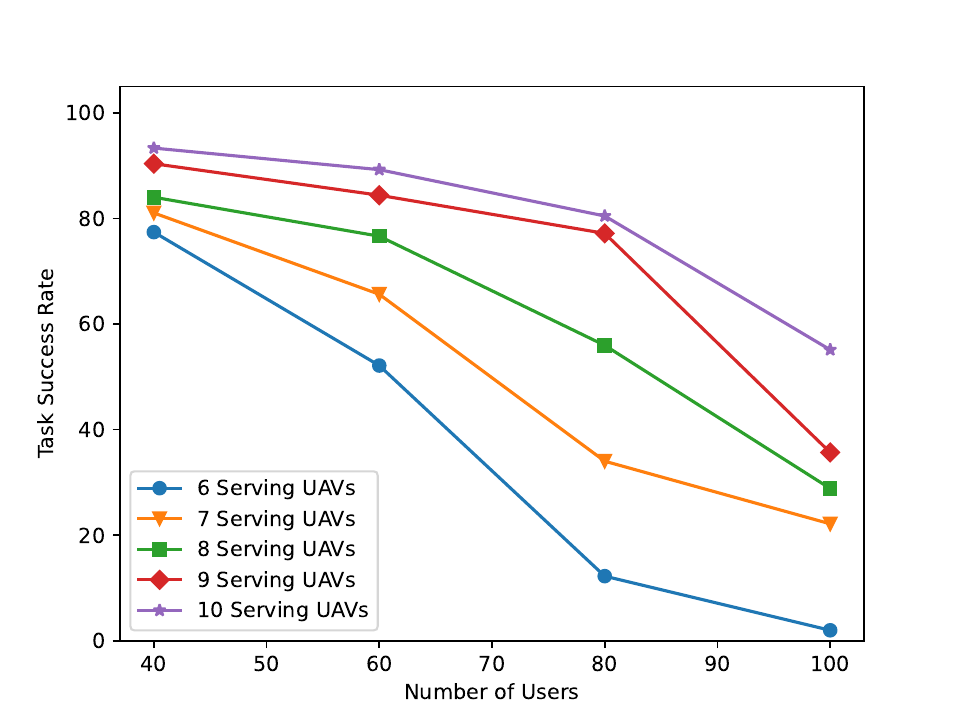}
\caption{Effect of the number of users and serving UAVs on DeepAir.}
\label{deepAir-res}
\end{figure}

\begin{figure}[t]
\centering
\includegraphics[scale=0.6]{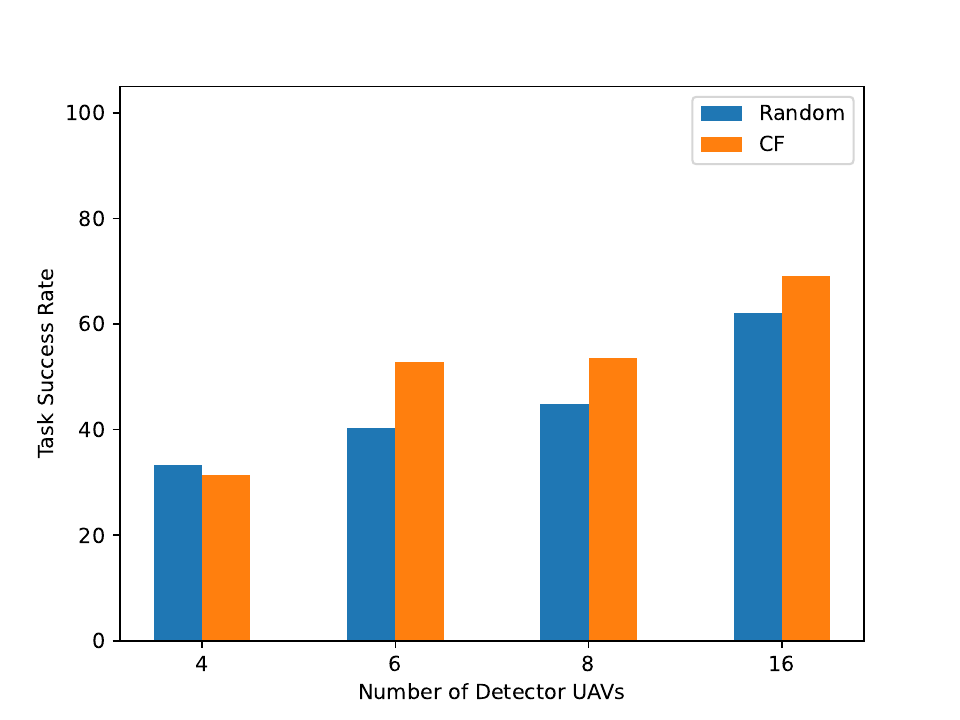}
\caption{Effect of the number of detector UAVs on benchmark methods using 80 users.}
\label{benchmark-exp}
\end{figure}

\subsection{Performance Experiments}
\label{experimentsPerf}

\begin{figure*}[!t]
\begin{subfigure}{0.50\textwidth}
\includegraphics[width=\linewidth]{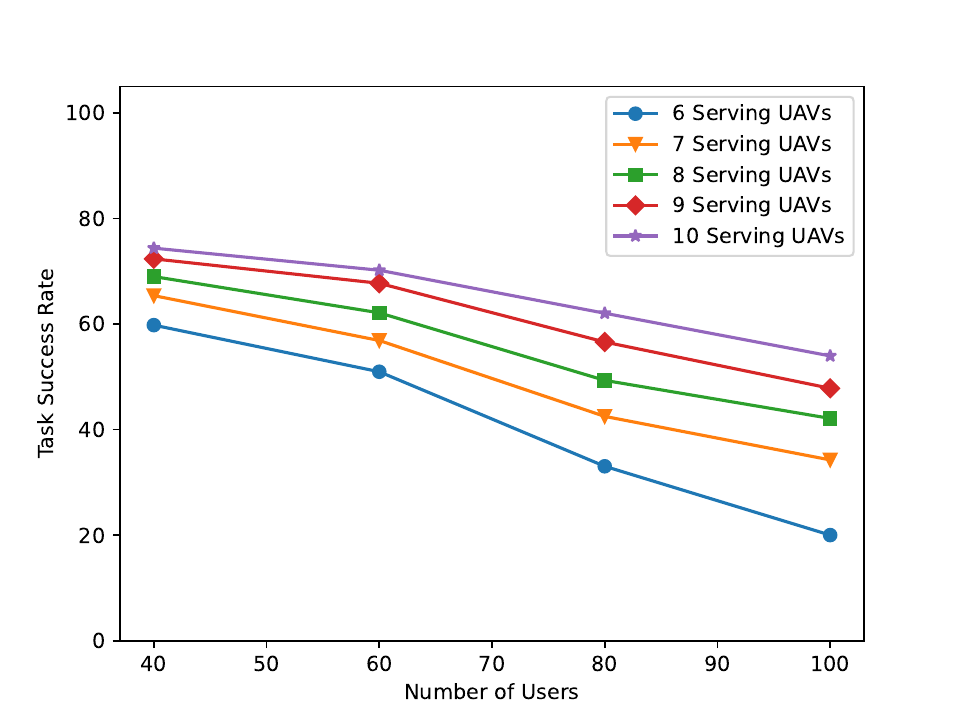}
\caption{Performance of Random-16} \label{Random-16}
\end{subfigure}
\hspace*{\fill} 
\begin{subfigure}{0.50\textwidth}
\includegraphics[width=\linewidth]{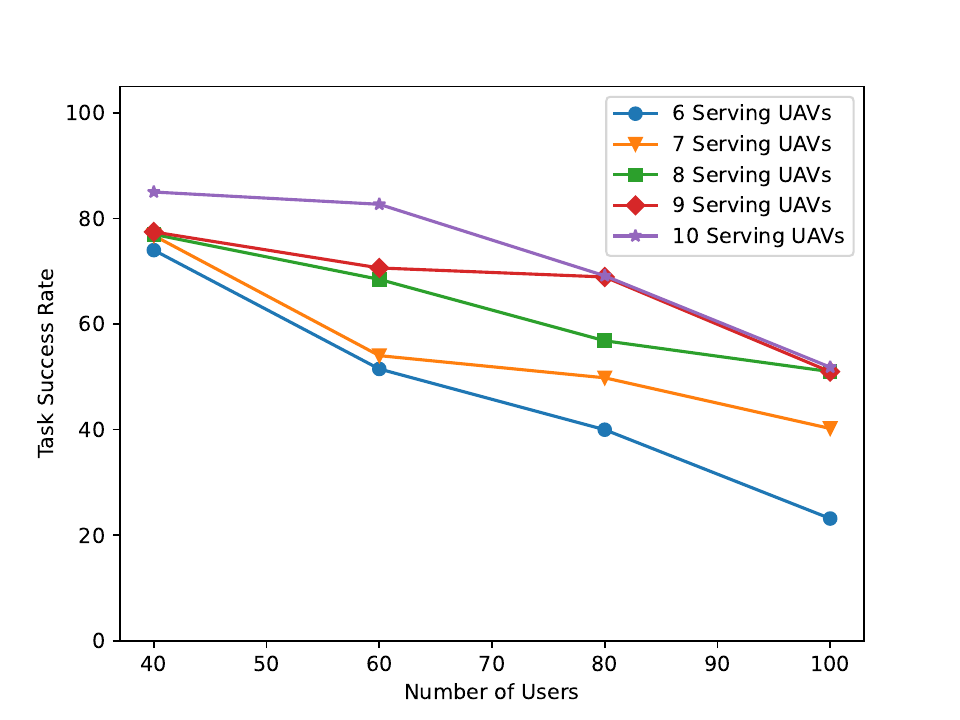}
\caption{Performance of CF-16} \label{CF-16}
\end{subfigure}
\caption{Results of benchmark methods}
\label{BenchmarkRes}
\end{figure*}

We first evaluate the performance of DeepAir considering the effect of varying numbers of users, and different serving UAVs. As shown in Figure \ref{deepAir-res}, the success rate of DeepAir is quite high based on the successfully placed detector UAVs through DRL. Note that based on the configuration in the experiments, at most six detector UAVs are used when we apply DeepAir. On the other hand, when the number of users increases based on the same number of attraction points, the task success rate for each serving UAV case decreases. This is an expected result since the computational capacity of those serving UAVs would not be sufficient to meet the higher number of tasks produced by each user in the environment. Similarly, a higher number of serving UAVs results in a higher task success rate since they provide more computational capacity.

Prior to the evaluation of the performance of benchmark methods, we first conducted experiments to observe their success rate using different numbers of detector UAVs. To this end, as shown in Figure \ref{benchmark-exp}, we compared 4, 6, 8, and 16 detector UAV cases using 80 users. As expected, using an increased number of detector UAVs provided a higher task success rate since the probability of covered users in the environment is higher in that case. Therefore, we used CF-16 and Random-16 as the benchmark methods in the experiments.

The performance of Random-16 and CF-16 methods based on different numbers of users and serving UAVs are shown in Figure \ref{BenchmarkRes}. The results manifest that CF-16 provides a better task success rate since it is a more structured approach. However, even though random decisions are taken in Random-16, when we use 9-10 serving UAVs, the task success rate is above 70\% which is quite good. On the other hand, when the number of users increases, the task success rate decreases in both methods for the same reasons that we elaborated on regarding the results of DeepAir.

\begin{figure}[t]
\centering
\includegraphics[scale=0.6]{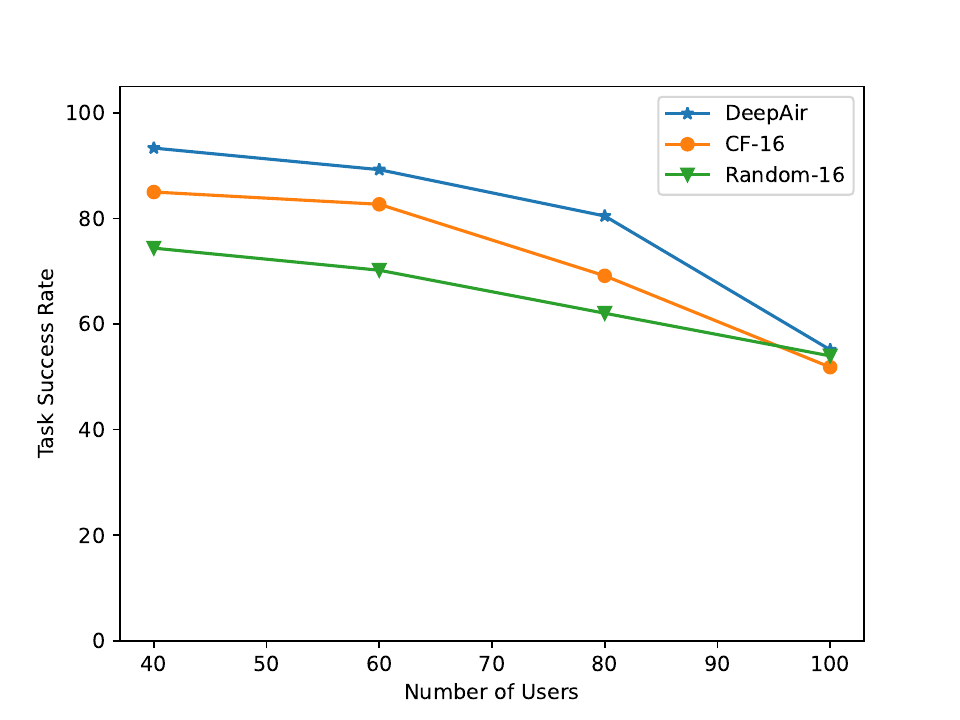}
\caption{Performance comparison between DeepAir and benchmark methods using 10 serving UAVs. While CF-16 and Random-16 use 16 detector UAVs, DeepAir outperform them using at most six detector UAVs.}
\label{comparedResults}
\end{figure}

After the evaluation of benchmark methods and DeepAir, next we compared their most successful deployments in which we used ten serving UAVs. Figure \ref{comparedResults} shows the results of this comparison. Based on the results, DeepAir outperforms both benchmark methods for different user counts. Note that both benchmark methods deployed 16 detector UAVs in this comparison, while DeepAir utilized at most six detector UAVs. On the other hand, the task success rate is close for each method when there are 100 users in the environment. This is the result of the insufficient computational capacity of serving UAVs deployed in the environment. Therefore, even if the best locations are selected for detector UAVs, the task success rate cannot exceed a certain value under those circumstances. Moreover, since the accuracy of location selection in Random-16 and CF-16 methods is less than that of DeepAir, the less number of connected users are served better using ten serving UAVs. This situation results in close values in terms of the task success rate when there are 100 users in the environment. In summary, although the user locations are successfully detected, if the number of serving UAVs is not sufficient for a certain load, the task success rate improvement requires more UAVs.

\section{Discussion}






Considering the fact that the unknown user location problem has not been deeply investigated in the literature, we think that several points should be discussed based on our observations and experiments throughout this study. 
We first noted about the discrete action space for each type of UAV. Since continuous horizontal actions would complicate the already complex problem regarding DRL, we applied a discrete action space. Moreover, and more importantly, applying a discrete action space alleviates the problem since it turns that into a maze problem in which there are several different prizes (RSSI power) in different sections of the environment. The agent learns to follow those small prizes to reach a bigger prize through episodes. Therefore, the convergence of the agents would be more quickly compared to the case of a continuous action space. As a result we could apply the DQN algorithm, which is a less complex regarding other DRL algorithms such as PPO, and DDPG.

Throughout our experiments, we also observed that higher RSSI due to a bigger number of users provides more accurate location prediction in DRL. As a result of this, more users can connect to the corresponding detector UAVs. Therefore, if the capacity of serving UAVs is so high, it is not so affected by the number of users, then the task success rate would be larger even though the load is increased. This is an important observation since, otherwise, the corresponding results would be evaluated incorrectly. For this reason, the selection of the capacity of serving UAVs and the required CPU cycles for user tasks are significant for manifesting the accuracy of the experiments.


\section{Conclusion}


In this study, we investigated the unknown user location problem in a UAV-assisted environment. The corresponding environment can be a disaster site, wilderness, or a rural area in which user devices cannot connect to any communication device and edge servers because of the lack of infrastructure. Moreover, each user device produces tasks that should be completed regarding their maximum tolerable delay which is not met by the computational capabilities of user devices. Therefore, those tasks should be offloaded to the related computational units. In order to achieve this in such an environment, sensing, localization, resource allocation, and  MEC capabilities should be provided together, sequentially. Therefore, we proposed DeepAir, a novel approach which uses DRL iteratively via detector UAVs that are responsible for sensing, localization, and resource allocation. Afterwards, MEC features are provided to those connected users by serving UAVs. Conducted experiments show that DeepAir provides a high task success rate by using a small number of detector UAVs in the environment regarding the benchmark methods.

In the future, we plan to take energy consumption into account since energy efficiency is crucial for the movements of the UAVs which would affect the performance. Therefore, we plan to optimize the trade-off between energy consumption and task success rate efficiently.

\bibliographystyle{IEEEtran}
\bibliography{DeepAir}

%

\begin{IEEEbiography}
[{\includegraphics[width=1in,height=1.25in,clip,keepaspectratio]{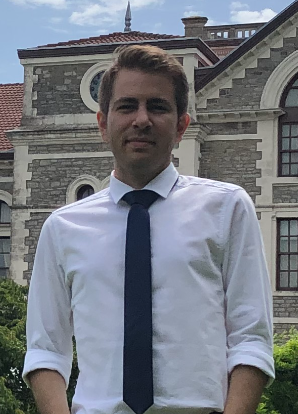}}]{Baris Yamansavacilar}
received his BS degree in Computer Engineering from Yildiz Technical University, Istanbul, in 2015. He received his MS degree in Computer Engineering from Bogazici University, Istanbul, in 2019. Currently, he is a PhD candidate in Computer Engineering Department at Bogazici University and a senior research engineer at Airties. His research interests include Machine Learning, Deep Reinforcement Learning, Edge/Cloud Computing, and Software-Defined Networks.

\end{IEEEbiography}

\begin{IEEEbiography}[{\includegraphics[width=1in,height=1.25in,clip,keepaspectratio]{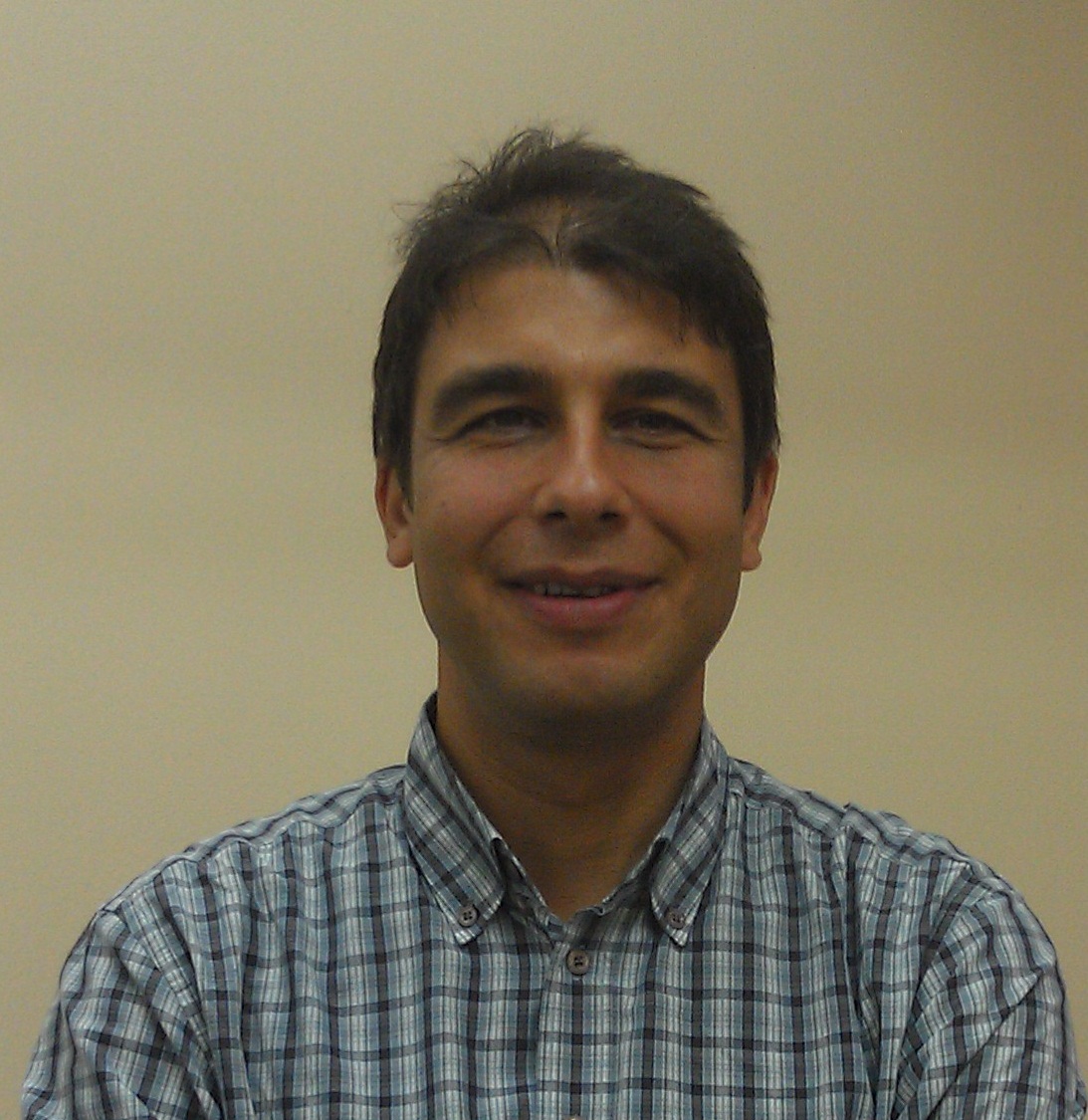}}]{Atay Ozgovde}
received the BS and MS degrees from Bogazici University, Istanbul, in 1995 and 1998, respectively. In 2002, he started working as a research assistant in the Computer Engineering Department, Bogazici University, where he completed the PhD degree in the NETLAB research group in 2009. He worked as a researcher at the WiSE-Ambient Intelligence Group to complete his postdoctoral research. Currently, he is an assistant professor in the Computer Engineering Department, Bogazici University. His research interests include wireless sensor networks, embedded systems, distributed systems, pervasive computing, SDN and mobile cloud computing. He is a member of the IEEE.
\end{IEEEbiography}

\begin{IEEEbiography}[{\includegraphics[width=1in,height=1.25in,clip,keepaspectratio]{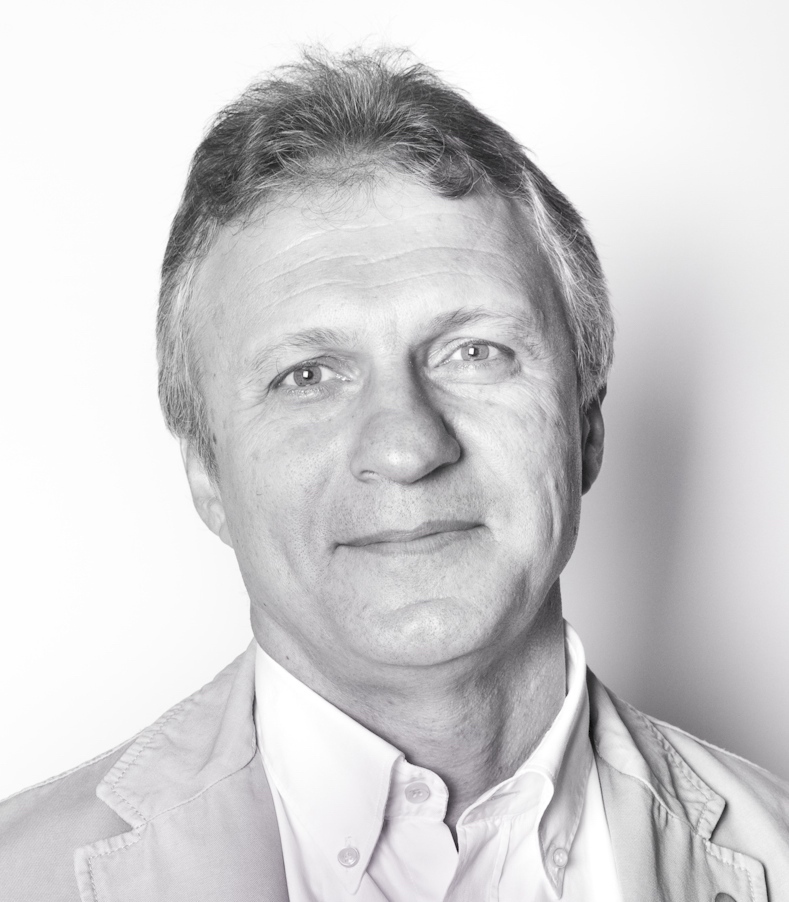}}]{Cem Ersoy}	
worked as an R\&D engineer in NETAS A.S. between 1984 and 1986. After receiving his PhD from Polytechnic University, New York in 1992, he became a professor of Computer Engineering in Bogazici University. Prof. Ersoy's research interests include wireless/sensor networks, activity recognition and ambient intelligence for pervasive health applications, green 5G and beyond networks, mobile cloud/edge/fog computing. Prof. Ersoy is a member of IFIP and was the chairman of the IEEE Communications Society Turkish Chapter for eight years.
\end{IEEEbiography}

\end{document}